\documentstyle{article}
\title{\bf Quintessence in Brane Cosmology}
\author{ Pedro F. Gonz\'{a}lez-D\'{\i}az.\\
Isaac Newton Institute, 20 Clarkson Road, Cambridge, CB3 0EH,
UK\\ and\\ Instituto de Matem\'{a}ticas y F\'{\i}sica Fundamental\\
Consejo Superior de Investigaciones Cient\'{\i}ficas\\ Serrano 121,
28006 Madrid, SPAIN\\ }
\date{March 3, 2000}
\begin{document}
\maketitle
\large
\setlength{\baselineskip}{0.5cm}

\begin{center}
{\bf Abstract}
\end{center}

In order to reconcile the non conventional character of brane
cosmology with standard Friedmann cosmology, we introduce in
this paper a slowly-varying quintessence scalar field in the
brane and analyse the cosmological solutions corresponding to
some equations of state for the scalar field. Different
compensation mechanisms between the cosmological constant in the
bulk and the constant tension resulting from the combined effect
of ordinary matter and the quintessence scalar field are derived
or assumed. It has been checked that the Randall-Sundrum
approach is not necessarily the best procedure to reconcile
brane and standard cosmologies, and that there exists at least
another compensating mechanism that reproduces a rather
conventional behaviour for an accelerating universe.

\pagebreak

In order to reconcile the mismatch of the scales of particle
physics and gravity it was recently suggested [1] that the
latter scale can be lowered all the way to the weak scale by
introducing large extra dimensions, so opening up the
possibility for new primordial cosmological scenarios. Randall
and Sundrum later proposed [2] that the size of such extra
dimensions could still be kept small, with the background metric
being a non flat slice of anti-de Sitter space due to the
existence of a negative bulk cosmological constant which is
exactly balanced by the tensions on the two branes occurring in
this scenario. It is the so-generated curved character of the
space-time which causes the physical scales on the two branes to
take on exponentially different values. This has prompted a lot
of activity [3] on the possibility that we live just in a
three-dimensional one-brane world embedded in a higher
dimensional space, while gravity pervades the whole highest
dimensional space which, contrary to the Kaluza-Klein spirit,
need not even be compact according to the Randall-Sundrum
philosophy. The cosmological evolution of the brane universe has
already been extensively investigated by many authors [4-8].

It was however shown by Bin\'{e}truy, Deffayet, Ellwanger and
Langlois [4] that the evolution of a brane-like universe is not
viable in the sense that the cosmological field equations
corresponding to it do not match the analogous
Friedmann-Robertson-Walker equations of standard cosmology, the
essential difference being that the energy density in the brane
appears quadratically, rather than linearly, in the right hand
side of the Einstein equations. It was later noted [5-7] that
brane and standard cosmologies could still be reconciled by
again using the Randall-Sundrum approach [2], so that a negative
cosmological constant is introduced in the bulk which is exactly
compensated by a constant tension in the brane in such a way
that the nonlinear term for the energy density in the field
equations would nearly vanish (leaving the usual
Friedmann-Robertson-Walker evolution) at late times, and is
relevant only at the earliest times, where cosmology becomes
highly non conventional. Thus, much as quantum effects are
currently thought to remarkably modify Einstein general
relativity only at the initial Planck era, one could also regard
the relevant primordial deviations from standard cosmology in
the brane to be caused by some sort of topological effects which
are only relevant in the realm of quantum gravity.

It is the aim of the present work to discuss and generalize the
above interpretation by obtaining particular exact solutions to
the five-dimensional Einstein equations for a brane universe
which corresponds to an observable ordinary matter in the brane
and a negative cosmological constant in the bulk that can also
now be compensated by the combined effect of the constant
ordinary-matter tension and a vacuum quintessence scalar-field
tension [9] in the brane. This cancellation mechanism so as the
one derived from the Randall-Sungrum condition, comes in our
model quite naturally from the constraint on the quintessence
potential which is derived from the field equations and
conservation laws (or as particular {\it ans\"atze} from that
potential in the case of constant scalar field). We note that
such a compensating mechanism actually generalizes the
Randall-Sungrum paradigm which in this paper will be extended to
encompass not just a vacuum constant brane tension, but also the
tension derived from the observable ordinary matter. Among the
cosmological models that we have found (which include
accelerating and decelerating open or closed universes with an
initial non conventional phase), there are some "exotic"
universes (that is universes with negative energy density for
ordinary matter) which correspond to particular solutions that
do not contain any non conventional initial evolution.

According to Bin\'{e}truy et al. [4,5], in a five-dimensional
space-time the generalized time-time component of the Friedmann
equations can be written on a three-brane as
\begin{equation}
\frac{\dot{R}^2}{R^2} =\frac{\kappa^2}{6}\rho_B
+\frac{\kappa^4}{36}\rho_b^2 +\frac{{\it C}}{R^4}-\frac{k}{R^2}
,
\end{equation}
where the overhead dot means derivative with respect to time,
$R$ is the scale factor in the brane, $\rho_B$ and $\rho_b$ are
the energy densities in the bulk and the brane, respectively,
${\it C}$ is an integration constant which is probably related
to the choice of the initial conditions for the universe and can
be interpreted as an effective radiation term [4,5], $k$ is the
topological curvature ($k=0,\pm 1$), and
\begin{equation}
\kappa^2=8\pi G_{(5)}\equiv M_{(5)}^{-3},
\end{equation}
with $G_{(5)}$ the five-dimensional Newton constant and $M{(5)}$
the five-dimensional reduced Planck mass. In what follows we
shall restrict ourselves to the flat case ($k=0$) and assume
that the boundary conditions for the universe are such that
${\it C}=0$. The nonlinear term proportional to $\rho_b^2$ in
Eq. (1) makes the cosmology resulting from this equation highly
non conventional. As pointed out above, several authors [3,5-7]
have suggested that the discrepancy between brane and standard
cosmologies may be greatly alleviated by introducing a constant
tension in the brane universe which compensates the negative
cosmological constant in the bulk, within the spirit of the
Randall-Sundrum approach. On the other hand, the conservation
law for the energy in the brane which is compatible with Eq. (1)
can be written as [5]
\begin{equation}
\dot{\rho}_b +3\frac{\dot{R}}{R}\left(\rho_b +p_b\right)=0 ,
\end{equation}
with $p_b$ the pressure in the brane. We shall furthermore take
$p_M=0$ for the state equation of the ordinary fluid in the
brane.

If, instead of a constant tension term, we introduce in the
brane a vacuum scalar quintessence field $\phi$, with state
equation [9]
\begin{equation}
p_{\phi}=\omega\rho_{\phi} ,\;\;\; 0\leq\omega<-1
\end{equation}
(with $\omega=-1$ corresponding to the constant tension case),
which behaves like a perfect fluid, then we have for the energy
density in the brane,
\begin{equation}
\rho_b=\rho_M+\rho_{\phi}
\end{equation}
(with $\rho_M$ the energy density for the ordinary matter), so
that Eq. (3) becomes
\begin{equation}
\dot{\rho}_b +3\frac{\dot{R}}{R}\left[\rho_M
+\rho_b(1+\omega)\right]=0 .
\end{equation}
The quintessence field $\phi$ is assumed to be a slowly varying
scalar field which can be defined by
\[\kappa^2\rho_{\phi}=\frac{1}{2}\dot{\phi}^2 +V(\phi)>0 ,\;\;
\kappa^2 p_{\phi}=\frac{1}{2}\dot{\phi}^2 -V(\phi)\leq 0 ,\]
where $V(\phi)$ is the potential energy for the field $\phi$.
Furthermore, since no interaction between the scalar field
$\phi$ and ordinary matter $M$ is assumed to occur in the brane,
one can take independent conservation laws for $M$ and $\phi$
[10]
\begin{equation}
\rho_M=\rho_{M0}\left(\frac{R_0}{R}\right)^{3} ,\;\;\;
\rho_{\phi}=\rho_{\phi
0}\left(\frac{R_0}{R}\right)^{3(1+\omega)} ,
\end{equation}
with the subscript $0$ denoting current value. Using then the
second of Eqs. (7), one can integrate Eq. (6) to yield
\begin{equation}
R^3\rho_b -\frac{R_0^{3(1+\omega)}\rho_{\phi 0}}{R^{3\omega}}=D
,
\end{equation}
where $D$ is an integration constant that gives the total mass
in the brane universe, $D\equiv M$. Replacing the expression for
$\rho_b$ obtained from Eq. (8) in the field equation (1) for
$k=0$ and ${\it C}=0$, we finally get a differential constraint
on the scale factor $R$ which must be satisfied by all possible
solutions:
\begin{equation}
\dot{R}^2 =\frac{\kappa^2}{6}\rho_{B}R^2
+\frac{\kappa^4}{36}\left[\frac{M}{R^2}+\frac{\rho_{\phi
0}R_0^{3(1+\omega)}}{R^{3\omega+2}}\right]^{2} .
\end{equation}

In order to set a suitable compensating mechanism able to
alleviate or solve the discrepancy between brane and standard
cosmologies and, in particular, to check whether the
Randall-Sundrum approach or other possible conditions can be
successfully applied with that purpose in the case $\omega=-1$,
we shall derive from Eq. (9) another constraint, that on the
scalar-field potential $V(\phi)$ satisfying the field equation
(1) and all conservation laws (7) and (8). From these
expressions and the definition of the field $\phi$, first one
obtains
\begin{equation}
\frac{R}{R_0}
=\left(\frac{V}{V_0}\right)^{-\frac{1}{3(1+\omega)}} ,\;\;\;
\frac{\dot{R}}{R_0 H_0} =
\left(\frac{V}{V_0}\right)^{-\frac{3\omega+5}{6(1+\omega)}}\frac{V'}{V_0
'} ,
\end{equation}
where $H_0$ is the current value of the Hubble constant,
$'=d/d\phi$, and [10]
\begin{equation}
V_0 =\frac{1-\omega}{2}\kappa^2\phi_{\phi 0} ,\;\; V_0
'=\pm\frac{3H_0(1-\omega)}{2}\sqrt{(1+\omega)\kappa^2\rho_{\phi
0}} .
\end{equation}
From expressions (10) and (11), Eq. (9) can now be directly
transformed into the wanted constraint. It is nevertheless
convenient first introducing the dimensionless cosmological
parameters
\begin{equation}
\Omega_B =\frac{\kappa^2 \rho_B}{3H_0^4} ,\;\; \Omega_M
=\frac{\kappa^2 \rho_{M0}}{3H_0^2} ,\;\;
\Omega_{\phi} =\frac{\kappa^2 \rho_{\phi 0}}{3H_0^2} .
\end{equation}
We then obtain for the constraint on the scalar potential:
\begin{equation}
\frac{V'^{2}}{H_0^2 V_0'^{2}}
=\frac{\Omega_B}{2}\left(\frac{V}{V_0}\right)
+\frac{1}{4}\left[\Omega_M\left(\frac{V}{V_0}\right)^{\frac{\omega+3}{2(1+\omega)}}
+\Omega_{\phi}\left(\frac{V}{V_0}\right)^{\frac{3}{2}}\right]^{2}
.
\end{equation}
This expression has been derived by implicitly assuming that
$\omega\neq -1$. For $\omega= -1$, $V=V_0\neq 0$ and
$\phi=\phi_0\neq 0$ are both constant, and $V_0 '=0$. Thus, in
the case that the field $\phi$ reduces to a cosmological
constant $\Lambda$ in the brane, we attain that the quantity
\[I\equiv\Omega_B+
\frac{1}{2}\left(\Omega_M+\Omega_{\Lambda}\right)^2 \]
(or $I=\Omega_B+1/2$ if we use the triangular cosmological
constraint $\Omega_k+\Omega_M+\Omega_{\Lambda}=1$ on the brane)
is mathematically indeterminate. One may then employ different
{\it ans\"atze} for the value of $I$. Three of such possible
conditions will appear to be particularly interesting. If we set
$I=\Omega_M^2/2$, the Randall-Sundrum condition $\Omega_B=-
\Omega_{\Lambda}^2/2$ is recovered, if we set $I=0$, it follows
that $\Omega_B=-1/2$, and finally if $I=1/2$, then $\Omega_B=0$.
All of these three ans\"atze will be used later on this paper.

From Eqs. (10) we can now derive the expression
\begin{equation}
\phi-\phi_0= -3(1+\omega)\frac{V_0 H_0}{V_0 '} \int
dt\left(\frac{R}{R_0}\right)^{-\frac{3}{2}(1+\omega)} ,
\end{equation}
which will be useful in what follows as well. For quintessence,
the most interesting models are those where $\omega$ is not a
constant and are defined in terms of an inverse-power law
potential for the field $\phi$ [11]. These models improve the
fine-tuning problem associated with the cosmological constant
[11], they solve the cosmic coincidence problem [12], and they
can be implimented in the realm of high energy physics [13].
Actually, one can obtain solutions to constraint (13) having the
form of an inverse-power law, for a generic $\omega$, only if
particular values for the cosmological parameters $\Omega$'s are
assumed (for example, if we set $\Omega_B=\Omega_M=0$, then the
potential that satisfies (13) has the form $V\propto\phi^{-2}$),
but one cannot obtain analytical solutions in closed form for
generic values of $\omega$ and the $\Omega$'s. Therefore, in the
remaining of this paper we shall consider different solutions to
the constraint equations on $V$ and $R$ for given particular
values of the quintessence parameter $\omega$ [9], and discuss
their physical motivation in the relevant cases. Let us start
with $\omega=0$ for which case the solution to constraint (9) is
\begin{equation}
R(t)=
R_0\left[\sinh\left(\sqrt{\frac{9H_0^2\Omega_B}{2}}t\right)\right]^{\frac{1}{3}}
.
\end{equation}
Similarly to how it happens for the solution that corresponds to
the case in which there are neither a cosmological constant nor
a quintessence field in the brane, the scale factor (15) clearly
has a nonconventional behaviour at early times, $R\propto
t^{1/3}$, followed by an exponential expansion at any late
times. Inserting the expression of $R$ given by Eq. (15) in the
general expression (14) and using Eqs. (11) for $\omega=0$, we
have
\begin{equation}
\phi-\phi_0= \pm\sqrt{\frac{2\Omega_{\phi}}{3\Omega_B}}
F\left[\arccos\left(\frac{1 -\sinh(a_0 t)}{1+\sinh(a_0
t)}\right), \frac{1}{\sqrt{2}}\right] ,
\end{equation}
where $F$ denotes the elliptic integral of the first kind [14],
and
\begin{equation}
a_0=\sqrt{\frac{9H_0^2\Omega_B}{2}} ;
\end{equation}
hence it follows
\begin{equation}
\sinh(a_0 t)= \frac{1- {\rm
cn}\left[\sqrt{\frac{3\Omega_B}{2\Omega_{\phi}}}(\phi-\phi_0)\right]}{1+
{\rm
cn}\left[\sqrt{\frac{3\Omega_B}{2\Omega_{\phi}}}(\phi-\phi_0)\right]}
,
\end{equation}
in which ${\rm cn}$ is a Jacobian elliptic function [15]. Using
then the first of the Eqs. (10) we obtain for the scalar
potential
\begin{equation}
V(\phi)=\frac{V_0}{\sinh(a_0 t)}= V_0\left\{ \frac{1+
{\rm
cn}\left[\sqrt{\frac{3\Omega_B}{2\Omega_{\phi}}}(\phi-\phi_0)\right]}{1-
{\rm
cn}\left[\sqrt{\frac{3\Omega_B}{2\Omega_{\phi}}}(\phi-\phi_0)\right]}\right\}
.
\end{equation}
It can be easily seen that the scalar potential (19) satisfies
constraint (13), provided $\Omega_M+\Omega_{\phi}=1$ and
$\Omega_B=1/2$. The latter condition implies a positive
cosmological constant in the bulk and corresponds to an ansatz
$I=1$, so that neither the Randall-Sungrum approach nor any of
the other two above-alluded conditions can hold in this case.
Clearly, if we set the initial conditions inmediately after
inflation, i.e. at a redshift $z\sim 10^{28}$, then
$\phi-\phi_0\propto F\left[\arccos(z/(z+2)),1/\sqrt{2}\right]$,
which is very small initially and along most of its evolution,
except when it approaches current values where $\phi-\phi_0\sim
1$, so providing a reason why the quintessential field starts
dominating only now, and hence solving the cosmic coincidence
problem [13]. On the other hand, for most of its cosmological
evolution, potential (19) can be approximated as $V+V_0\propto
V_0(\phi-\phi_0)^{-2}$, i.e. an inverse-power law potential
which might be linked to particle physics models [13].

We consider next the ansatz $I=1/2$ (i.e. $\Omega_B=0$) for the
case that the quintessence state equation takes the particular
expression corresponding to $\omega=-1/2$. We distinguish the
approximate early and late time solutions:
\begin{equation}
R(t)= R_0\left(\frac{3}{2}\Omega_M H_0 t\right)^{\frac{1}{3}}
\end{equation}
for early times, and
\begin{equation}
R(t)= R_0\left(\frac{3}{4}\Omega_{\phi}H_0
t\right)^{\frac{2}{3}}
\end{equation}
for late times. It is worth noticing that the same qualitative
behaviour for the scale factor as that given by solutions (20)
and (21) (that is, an initial non conventional expansion,
followed by a conventional one up to arbitrarily large time) was
also obtained using the Randall-Sungrum approach with a
cosmological constant in the brane [5]. Following then the same
steps as for $\omega=0$, we finally get again in this case an
inverse-power law for the scalar potential; i.e.
\begin{equation}
V(\phi)\propto\left(\phi-\phi_0\right)^{-2} ,
\end{equation}
with the proportionality constants being given by simple
functions of $\Omega_{\phi}$ and $\Omega_M$, both for early and
late times. It can be checked that this potential satisfies
constraint (13) by simply imposing the cosmological triangular
condition $\Omega_k+\Omega_{\phi}+\Omega_M=1$ in the brane.
Potential (22) may once again be implemented in the realm of
high energy physics and from it and the first of Eqs. (10), one
can deduce that the quintessential field becomes in this case
proportional to $(1+z)^{-3/4}$, so that one can also solve the
cosmic coincidence problem as well.

We finally come to the case $\omega=-1$. As pointed out above,
this corresponds to having a cosmological constant $\Lambda$ in
the brane, so that $V=V_0$, $\phi=\phi_0$ and $I$ becomes
indeterminate. Three ans\"atze will be considered for $I$:
$I=\Omega_M^2/2$ (corresponding to the Randall-Sungrum
approach), $I=0$ and $I=1/2$. These particular approaches are
associated with values of the cosmological constant in the bulk
given by $\Omega_B=-\Omega_{\Lambda}^2/2$, $\Omega_B=-1/2$ and
$\Omega_B=0$, respectively. For the first of these conditions,
the solution reads:
\begin{equation}
R(t)= R_0\left[\frac{9H_0^4
\Omega_M\Omega_{\Lambda}}{8}\left(t^2
+\frac{4t}{3H_0^2\Omega_{\Lambda}}\right)\right]^{\frac{1}{3}} ,
\end{equation}
which corresponds to the same qualitative behaviour as for the
solution obtained when we take $I=1/2$, $\omega=-1/2$, such as
it was mentioned above. If we allow the quantity $I$ to be in
the close neighbourhood of the Randall-Sundrum value
$\Omega_M^2/2$ and keep $\ell =I-\Omega_M^2/2$ nonzero but very
small, then it is obtained for $R$
\begin{equation}
R(t)= R_0\left\{\frac{1}{2\ell^2\kappa^4}\left[\left(R_0^{-3}
-\frac{\Omega_M\Omega_{\Lambda}}{2}\right)\sinh\left(3\ell\kappa^2
t\right) +
\frac{\Omega_M\Omega_{\Lambda}}{2}\left(\cosh\left(3\ell\kappa^2
t\right)-1\right)\right]\right\}^{\frac{1}{3}} .
\end{equation}
Solution (24) describes the evolution of a brane universe which
initially expands non conventionally, $R\propto t^{1/3}$, to
enter then the customary regime, $R\propto t^{2/3}$, and finally
an exponential realm, $R\propto\exp(3\ell\kappa^2 t)$, that keep
holding forever. We thus recover the solution first derived by
Bin\'{e}truy et al. [5] in the case of a general equation of state
for the ordinary matter.

The use of the ans\"atz $\Omega_B=0$ leads to a solution to the
constraint on $R$ which reads:
\begin{equation}
R(t)=
R_0\left\{\frac{\Omega_M}{\Omega_{\Lambda}}\left[\exp\left(
\frac{\kappa^2\Omega_{\Lambda}}{R_0^3\Omega_M}t\right)-
1\right]\right\}^{\frac{1}{3}} ,
\end{equation}
which, for a convenient choice of the cosmological parameters
$\Omega_M$ and $\Omega_{\Lambda}$, shows a qualitative behaviour
similar to that is predicted by solution (24).

Let us finally consider the case where $I=0$, so that
$\Omega_B=-1/2$. Then from the constraint on $R$ we have for the
cosmological time
\begin{equation}
t= -\int\frac{dx}{x\sqrt{M^2x^2 +2M\rho_{\Lambda}x
+\left(\rho_{\Lambda}^2-\frac{9H_0^4}{\kappa^4}\right)}} ,
\end{equation}
where $x=R^{-3}$. Now, we shall examine the three possible
situations which appear depending on whether
$\Omega_{\Lambda}^2$ is larger, equal or smaller than unity. If
$\Omega_{\Lambda}^2>1$ then we get from Eq. (26):
\[ R(t)=
\frac{1}{\left(\Omega_{\Lambda}^2-1\right)^{\frac{1}{3}}}
\left\{\left(M^2-\frac{\kappa^4}{36H_0^4}\right)
\sinh\left(\frac{3}{2}\sqrt{\Omega_{\Lambda}^2-1}H_0^2
t\right)\right.\]
\begin{equation}
\left.+\left(M^2+\frac{\kappa^4}{36H_0^4}\right)
\left[\cosh\left(\frac{3}{2}\sqrt{\Omega_{\Lambda}^2-1}H_0^2
t\right)-1\right]\right\}^{\frac{1}{3}} .
\end{equation}
Now, if $M>\kappa^2/6H_0^2$ one should expect from solution (27)
the same qualitative evolutive behaviour as that was obtained
from solution (24), including the initial non conventional
phase. More interesting are the cases where the mass of the
brane universe is constrained to be $M=\pm\kappa^2/6H_0^2$. Then
from the cosmological triangular condition on the brane we have
$\Omega_{\Lambda}=1\mp\kappa^2/(18H_0^4 R_0^3)$. In these cases
the scale factor (27) reduces to
\begin{equation}
R(t)=\left\{\frac{2M^2}{\Omega_{\Lambda}^2-
1}\left[\cosh\left(\frac{3}{2}\sqrt{\Omega_{\Lambda}^2-1}H_0^2
t\right)-1\right]\right\}^{\frac{1}{3}} ,
\end{equation}
which represents a universe which initially expands according to
conventional cosmology ($R\propto t^{2/3}$) to finally grow
exponentially, without passing through any non conventional
$t^{1/3}$-phase. Thus, if it turned finally out that the
observable universe is actually accelerating [16], the brane
solution (28) could be regarded as a good candidate to describe
it. The price to be paid for getting such a conclusion is to
allow for an universe with either a large positive mass
$\Omega_M>2$ or a negative energy density $\Omega_M<0$.

If $\Omega_{\Lambda}^2<1$, then we obtain the closed solution
\begin{equation}
\frac{R(t)}{R_0}=
\left\{\frac{\Omega_M}{1-\Omega_{\Lambda}^2}
\left[\Omega_{\Lambda}
+\sin\left(\frac{3}{2}h_0^2\sqrt{1-
\Omega_{\Lambda}^2}t\right)\right]\right\}^{\frac{1}{3}}
-\left(\frac{\Omega_M\Omega_{\Lambda}}{1-
\Omega_{\Lambda}^2}\right)^{\frac{1}{3}} ,
\end{equation}
which appears to be not quite fashionable not just for its
initial non conventional behaviour, but mainly for its
disconform to recent cosmological observations. More interesting
appears to be the solution that corresponds to the case
$\Omega_{\Lambda}^2=1$ which exactly coincides with the solution
obtained when one imposes the Randall-Sungrum approach (Eq.
(23)).

Before closing up, it seems interesting to notice an additional
physical motivation for the general model used in this letter.
If we include among the nongeometrical contributions to the
energy density entering the definition of the luminosity
distance $D_L$ [17] the contribution from the bulk $\Omega_B$,
then using the definition of the redshift in terms of the scale
factor and the first of Eqs. (10) we obtain for our flat model:
\begin{equation}
D_L\equiv D_L(\Omega_B)= \frac{V_0 '
(1+z)}{2H_0(\omega+1)V_0^{\frac{1}{2}+\frac{1}{3(\omega+1)}}}
\int{\phi(0)}^{\phi(z)}\frac{d\phi}{V^{\frac{3\omega+1}{6(\omega+1)}}}
,
\end{equation}
with $H_0$ the Hubble constant. Bringing then this expression
into the magnitude($m^{eff}$)-redshift($z$) relation [16], our
predictions can be compared with the results obtained in
observations of distant supernovas. For inverse-power law
potentials $V\propto\phi^{-\alpha}$, with $\alpha >0$, the use
of Eq. (30) leads to magnitude-redshift plots which predict a
suitable accelerating behaviour. An interesting physical
consequence is that the shape of these plots is insensitive to
the value of $\Omega_B$, even though $m^{eff}$ depends on
$\Omega_B$. It follows that one cannot extract any information
about the bulk from supernova observations.

Clearly, there appear to be many other interesting solutions
corresponding to other choices of quintessence parameter
$\omega$ and/or the indeterminate quantity $I$ when $\omega=-1$
which may be dealt with and interpreted following lines
analogous to those considered in this paper. Moreover, one
should check the stability of our $\omega$-constant potentials
to quantum corrections. However, the results obtained so far
seem to be forcefull enough to draw off the following
conclusion. When applied to brane cosmology, the Randall-Sungrum
approach is nothing but just another more condition among a
presumably large number of similar ans\"atze which can all
alleviate or even solve the mismatch between brane and standard
cosmology. It appears e.g. that for the case $\omega=-1$ there
actually exists at least a particular condition, other than that
of Randall and Sungrum, that is expressed as $\Omega_B=-1/2$,
which may even lead to a cosmological model without any non
conventional expansion phase. In one case this solution
corresponds to a universe filled with matter having negative
energy density. This situation would mean violation of the
classical energy conditions [18] and could imply the existence
of causality-violating processes involving superluminal travels
or closed timelike curves [19].

\vspace{.8cm}

\noindent {\bf Acknowledgements}

The author thanks C.L. Sig\"uenza for useful comments and a
careful reading of the manuscript. This work was supported by
DGICYT under Research Project No. PB97-1218.

\noindent\section*{References}
\begin{description}
\item [1] N. Arkani-Hamed, S. Dimopoulos, and G. Dvali, Phys.
Lett. B429 (1998) 263; I. Antoniadis, N. Arkani-Hamed, S.
Dimopoulos and G. Dvali, Phys. Lett. B436 (1998) 257.
\item [2] L. Randall and R. Sundrum, Phys. Rev. Lett. 83 (1999)
3370; 4690.
\item [3] C. Cs\'{a}ki, M. Graesser, L. Randall, and J. Terning,
{\it Cosmology of Brane Models with Rodion Stabilization},
hep-ph/9911406, and references therein.
\item [4] P. Bin\'{e}truy, C. Deffayet, and D. Langlois, {\it
Non-Conventional Cosmology from a Brane Universe},
hep-th/9905012.
\item [5] P. Bin\'{e}truy, C. Deffayet, U. Ellwanger, and D.
Langlois, {\it Brane Cosmological Evolution in a Bulk with
Cosmological Constant}, hep-th/9910219.
\item [6] C. Cs\'{a}ki, M. Graessner, C. Kolda, and J. Terning,
Phys. Lett. B426 (1999) 34.
\item [7] J.M. Cline, C. Grosjean, and G. Servant, Phys. Rev.
Lett. 83 (1999) 4245.
\item [8] K. Koyama and J. Soda, {\it Birth of the Brane
Universe}, gr-qc/0001033; J. Garriga and M. Sasaki, {\it
Brane-World Creation and Black Holes}, hep-th/9912118; {\it
Gravity in the Brane-World}, hep-th/9911055.
\item [9] R.R. Caldwell, R. Dare, and P.J. Steinhardt, Phys.
Rev. Lett. 80 (1998) 1582.
\item [10] E. Di Pietro and J. Demaret, {\it Quintessence: Back
to Basis}, gr-qc/9908071.
\item [11] I. Zatlev, L. Wang and P.J. Steinhardt, Phys. Rev.
Lett. 82 (1999) 896.
\item [12] P.J. Steinhardt, in {\it Critical Problems in
Physics}, edited by V.L. Fitch and D.R. Marlow (Princeton Univ.
Press, Princeton, USA, 1997).
\item [13] A. Masiero, M. Pietroni and E. Rocati, Phys. Rev. D61
(2000) 023504; K. Choi, {\it String or M Theory Axion as a
Quintessence}, hep-th/9902292; P.Brax and J. Martin, {\it The
Robustness of Quintessence}, astro-ph/9912046.
\item [14] I.S. Gradshteyn and I.M. Ryzhik, {\it Table of Integrals,
Series, and Products}, edited by A. Jeffrey (Academic Press, San
Diego, USA, 1994).
\item [15] M. Abramowitz and I. Stegun, {\it Handbook of
Mathematical Functions} (Dover, New York, USA, 1964).
\item [16] S. Perlmutter {\it et al.}, ApJ. 517 (1999) 565; A.G.
Riess {\it et al.}, Astron. J. 116 (1998) 1009.
\item [17] B.P. Schmidth {\it et al.}, ApJ. 509 (1998) 54.
\item [18] S.W. Hawking and G.F.R. Ellis, {\it The Large Scale
Structure of Space-Time} (Cambridge University Press, Cambridge,
UK, 1973).
\item [19] M. Visser, {\it Lorentzian Wormholes} (AIP, Woodbury,
New York, USA, 1996).

\end{description}

\end{document}